\documentclass[%
 aip,
 amsmath,amssymb,
 reprint,%
]{revtex4-1}

\usepackage{graphicx}
\usepackage{dcolumn}
\usepackage{bm}

\usepackage[utf8]{inputenc}
\usepackage[T1]{fontenc}
\usepackage{mathptmx}
\usepackage{textcomp}

\begin{document}

\preprint{AIP/123-QED}

\title{Ultrafast generation and detection of propagating coherent acoustic phonon wave packets in ultra-thin iron pnictide films}

\author{D. Cheng, B. Song}
\affiliation{Department of Physics and Astronomy, Iowa State University, and Ames Laboratory, Ames, IA 50011 USA
}%
\author{J.~H.~Kang}
\author{C.~Sundahl}
\affiliation{%
	Department of Materials Science and Engineering, University of Wisconsin-Madison, Madison, WI 53706, USA
}%
\author{L. Luo, J-M. Park}
\affiliation{Department of Physics and Astronomy, Iowa State University, and Ames Laboratory, Ames, IA 50011 USA
}%
\author{Y. G. Collantes}
\author{E. E. Hellstrom}
\affiliation{%
Applied Superconductivity Center, National High Magnetic Field Laboratory, Florida State University, Tallahassee, FL 32310, USA
}%
\author{M.~Mootz}
\author{I.~E.~Perakis}
\affiliation{%
Department of Physics, University of Alabama at Birmingham, Birmingham, AL 35294-1170, USA
}%
\author{C.~B.~Eom}
\affiliation{%
	Department of Materials Science and Engineering, University of Wisconsin-Madison, Madison, WI 53706, USA
}%
\author{J. Wang}%
\email{jgwang@iastate.edu; jgwang@ameslab.edu}
\affiliation{Department of Physics and Astronomy, Iowa State University, and Ames Laboratory, Ames, IA 50011 USA
}%

\date{\today}

\begin{abstract}
We observe pronounced oscillations in differential reflectivity of 9 nm and 60 nm BaFe\textsubscript{2}As\textsubscript{2} (Ba-122) thin films using ultrafast optical spectroscopy. Our studies show that the oscillations result from propagating longitudinal acoustic (LA) phonon wave packets with strong thickness and temperature dependence. Particularly, the experimentally measured oscillation frequency approaches to 50 GHz for the ultra-thin film. 
Our calculations show that Young's modulus of 9 nm thin film is nearly four times as large as that of 60 nm thin film, consistent with the experiment. The increase in Young's modulus as thickness decrease was attributed to the decrease in parent Ba-122 tetragonality $c/a$ near the film-substrate interface due to material-substrate mismatch effect. 
The temperature-dependent change in LA phonon frequency was attributed to the change in parent Ba-122 othorhombicity $(a-b)/(a+b)$. 
\end{abstract}

\maketitle

%


The high-frequency phonons carry key information about the materials' elastic properties and were commonly detected by superconducting tunnel junctions and heat pulses\cite{review}. Both measurements were performed at low temperatures (usually in superconductivity regimes). Thanks to the development of ultrafast optical techniques, the modern approach is based on ultrafast lasers, which can generate and detect acoustic phonon wave packets, as seen in the pioneering work in 1984 by Thomsen \textit {et al}\cite{firstexp}. After the early attempts, enduring efforts have been devoted to generate and study phonon modes using ultrafast lasers\cite{iron, ref1, ref2, ref4, ref5}. The intense ultrafast optical pulse (pump) disturbs the thin film and generates acoustic phonons, which will bounce back and forth between film surfaces and cause a periodic modulation of optical properties. The modulation is subsequently measured by a time-delayed light pulse (probe) at sample's surface. The detected phonon mode is usually longitudinal acoustic (LA) waves since the transverse acoustic (TA) waves, or shear waves, could only be generated through anisotropic thermal expansion in thermally anisotropic materials\cite{shearmode}. Unlike superconducting tunnel junctions and heat pulses techniques, the ultrafast laser method can also be used at much higher temperatures\cite{thermal,a, b, c, d, e}. 

The ultrafast optical technique expands our knowledge beyond bulks, with the unprecedented sensitivity, time resolution and applicability in ultra-thin films even down to $\sim$10 nm. Specifically, we can examine the elastic properties, like Young's modulus, of ultra-thin films compared with its bulk counterpart. The properties might have been significantly altered due to the strain effect and lower mobility of dislocation. Such studies and pertinent discussions \cite{Young, cc} have been rare, especially for ultra-thin film down to 10nm and temperature dependence of Young's modulus, which has motivated the present study on BaFe$_2$As$_2$ (Ba-122) films.

The Ba-122 phase of iron pnictide materials hosts a series of layered intermetallics \cite{HC, 122_1, 122_2, 122_3, 122_4, 122_5} and their derivatives \cite{bqs}, subject to a tetragonal-orthohombic transition and a phase diagram involving magnetism and superconductivity \cite{David}. It has been shown that superconductivity of ultra-thin film Ba-122 could be enhanced by tuning the othorohmbicity $(a-b)/(a+b)$ of the unit cell\cite{samplegrower}. Change of  sample's othorohmbicity would yield the change of elastic properties like intrinsic strain, indicative of the fact that the Young's modulus of ultra-thin film used for the study should differ from that of the bulk. This makes the system a good candidate to study the anomalous elastic properties in thin films.

In this Letter, we demonstrate a powerful and versatile method to measure Young's modulus of Ba-122 ultra-thin films through coherent LA phonon spectroscopy. This is implemented by ultrafast two-color differential reflectivity spectroscopy utilizing the pump (800 nm, 1.55eV) and probe (400 nm, 3.1 eV) laser pulses. Two thin films of 9 nm and 60 nm are measured and compared. We also performed temperature-dependent phonon frequency measurement to study its connections to the structural phase transition. We found that Young's modulus of Ba-122 film increase with the decrease in film thickness, which was mainly attributed to the material-substrate mismatch effect, i.e., tetragonality $c/a$ change near the film-substrate interface. The temperature-dependent change in LA phonon frequency was attributed to the change in parent Ba-122 othorhombicity $(a-b)/(a+b)$. 

Parent Ba-122 thin films were grown on single-crystal cubic substrates by pulsed laser deposition with a KrF (248 nm) ultraviolet excimer laser at 650-740 \textdegree{}C. High-quality epitaxial films were synthesized on lithium fluoride (LiF). The 800 nm laser is directly from a Ti:Sa amplifier with 40 fs pulse duration and 1 kHz repetition rate, which is used as pump to photoexcite the samples out of equilibrium. The 400 nm laser is frequency-doubled from the 800 nm beam by using a Beta Barium Borate (BBO) crystal, which is used as a probe to detect the transient reflectivity change of the samples. The pump-probe time delay is controlled by an optical delay stage. Both pump and probe beams are at near normal incidence of the samples.

Figures 1a and 1b illustrate the generation and detection methods of LA in Ba-122 thin films and the crystal structure of bulk Ba-122 and LiF. Figure 1c shows the schematics of $a$-$c$ over layers for Ba-122 thin film. The lattice constants are different at the top and the bottom (in contact with substrate) due to the intrinsic stress, more specifically the material-substrate mismatch. The lattice constants near the top are $a=3.9612$ \AA, and $c= 13.0061$ \AA. The lattice constant \textit{a} of Ba-122 single crystal is smaller than that of LiF substrate (4.027 \AA), which causes larger lattice constant $a^\prime$  and smaller lattice constant $c^\prime$ (thus smaller tetragonality $c^\prime$/$a^\prime$) near the bottom. Generally, the larger the number of layers, the less the material-substrate mismatch effect.

\begin{figure*}
	\includegraphics[width=\textwidth]{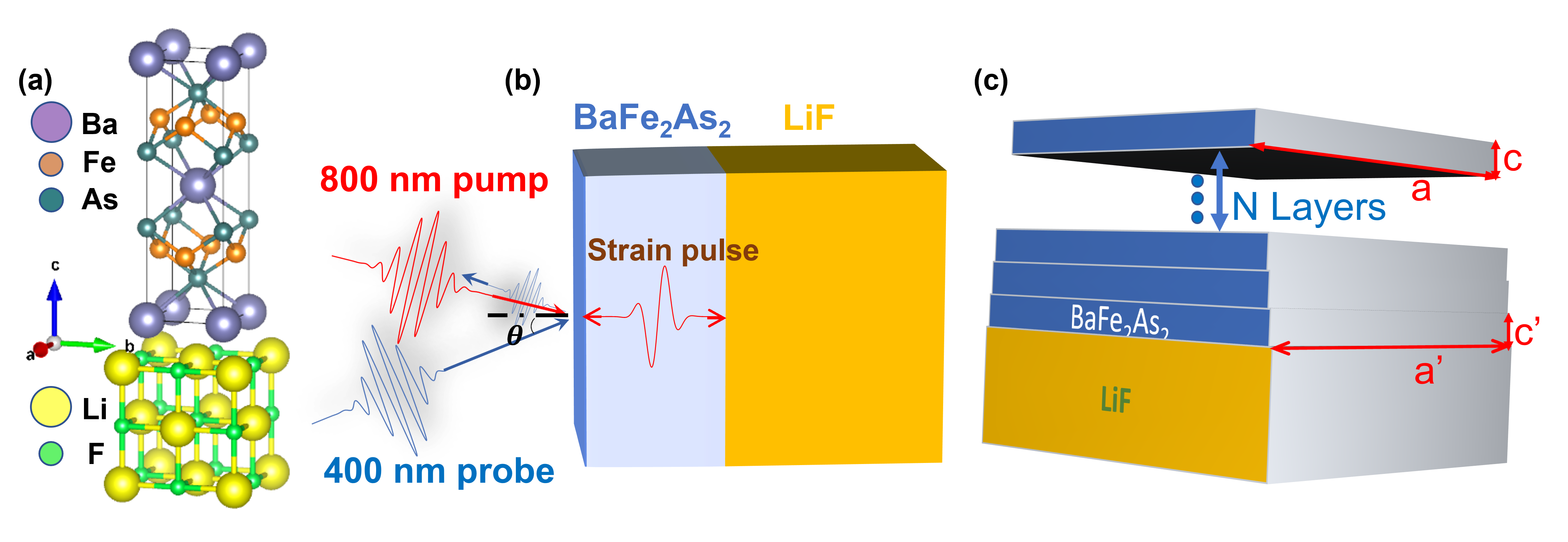}
	\caption{(a) Crystal structures of bulk Ba-122 and LiF\cite{crystalstructure}. (b) Generation and detection of coherent acoustic phonons in Ba-122 thin films. The coherent phonons generated by the 800 nm pump pulse can be detected by the time-delayed 400 nm probe pulse through the photoelastic effect.  (c) Schematics of $a$-$c$ over layers for Ba-122 thin film.}
	\label{fig1} 
\end{figure*}

Figures 2a and 2b show the transient reflectivity $\Delta R/R$ of 9 nm and 60 nm Ba-122 thin films, respectively, at $T$=5 K. Besides exponential decay of $\Delta R/R$ signal, periodic oscillations in $\Delta R/R$ were observed in both samples. The acoustic pulse generated at the surface propagates through the oscillations of thin film's sensitivity function and leads to the observed oscillations. We did theoretical fitting of the $\Delta R/R$ data and found that they cannot be fitted with a single exponential decay and that a biexponential decay works pretty well, which can be described as $\Delta R/R=R\textsubscript{0}+A\textsubscript{1}\text{exp}^{-(t-t_{0})/\tau\textsubscript{1}}+A\textsubscript{2}\text{exp}^{-(t-t_{0})/\tau\textsubscript{2}}$, where $R\textsubscript{0}$ is a time independent constant, $A\textsubscript{1}$ and $A\textsubscript{2}$ are the amplitudes of each exponential term, $t$ is the time delay, and $\tau\textsubscript{1}$ and $\tau\textsubscript{2}$ are the decay time for the first and second exponential terms, respectively.  According to our fitting,  $\tau\textsubscript{1}$ is 25 ps (17 ps) and $\tau\textsubscript{2}$ is 1 ns (2 ns) for 9 nm (60 nm) thin film, respectively. After subtracting the fitted biexponential decay background from $\Delta R/R$ signal, we performed Fourier transform of the residual oscillation signal, which are shown in Fig. 2c and 2d, respectively. For the 9 nm film, there is one peak at $\sim$47.5 GHz (Fig. 2c) and for the 60 nm film,  the peak is at $\sim$25 GHz (Fig. 2d). We attribute the 47.5 GHz and 25 GHz resonances to the coherent LA phonon modes generated by the pump beam since TA can only be generated in materials with anisotropic thermal expansion coefficients.

The acoustic phonon mode resonance \textit{f} is related to the sound velocity and probe wavelength through $f=2nu\text{cos}(\theta)/\lambda$  \textsuperscript{(Ref. 10)}, with $\theta$ being the angle of incidence of the probe beam (in our experiment the $\theta$ is close to zero) and $\lambda$ being the probe wavelength (400 nm). The refractive index $n$ for Ba-122 at 400 nm (25000 cm\textsuperscript{-1}) is 3.875. From \textit{f} which has been measured in Fig. 2c and 2d, we can calculate the sound velocity \textit{u}, which is 2456.77 m/s for the 9 nm film and is 1290.32 m/s for the 60 nm film. The sound velocity of the 60 nm film is nearly half of that of the 9 nm film and it is related to Young’s modulus through the equation $u=\sqrt{Y/\rho}$, with $Y$ being the Young’s modulus of the sample and $\rho$ the density of the sample. The density of the two thin films should be very similar. Therefore, the main difference in calculated sound velocity must be from the difference in Young’s modulus of the two samples. At fixed density $\rho$, $Y\propto u^2$, thus Young's modulus of 9 nm Ba-122 thin film is nearly four times as large as that of 60 nm thin film. The cause will be discussed later. 

We also measured the LA phonon mode resonance as a function of temperature. Figure 2e shows the transient reflectivity $\Delta R/R$ of the 9 nm film at various temperatures and Fig. 2f plots the LA phonon mode resonance as a function of temperature, extracted from Figure 2e. As seen from Fig. 2f, the  LA phonon mode resonance is nearly unchanged from 5 K to 50 K. At 80 K, the resonance starts to drop. This trend is perfectly correlated with temperature-dependent othorhombicity of the 9 nm Ba-122 sample\cite{samplegrower}. Therefore, we attribute the decrease in the LA mode resonance to the decrease in Young’s modulus originating from the decrease of orthorhombicity $(a-b)/(a+b)$.

\begin{figure*}
	\includegraphics[width=\textwidth]{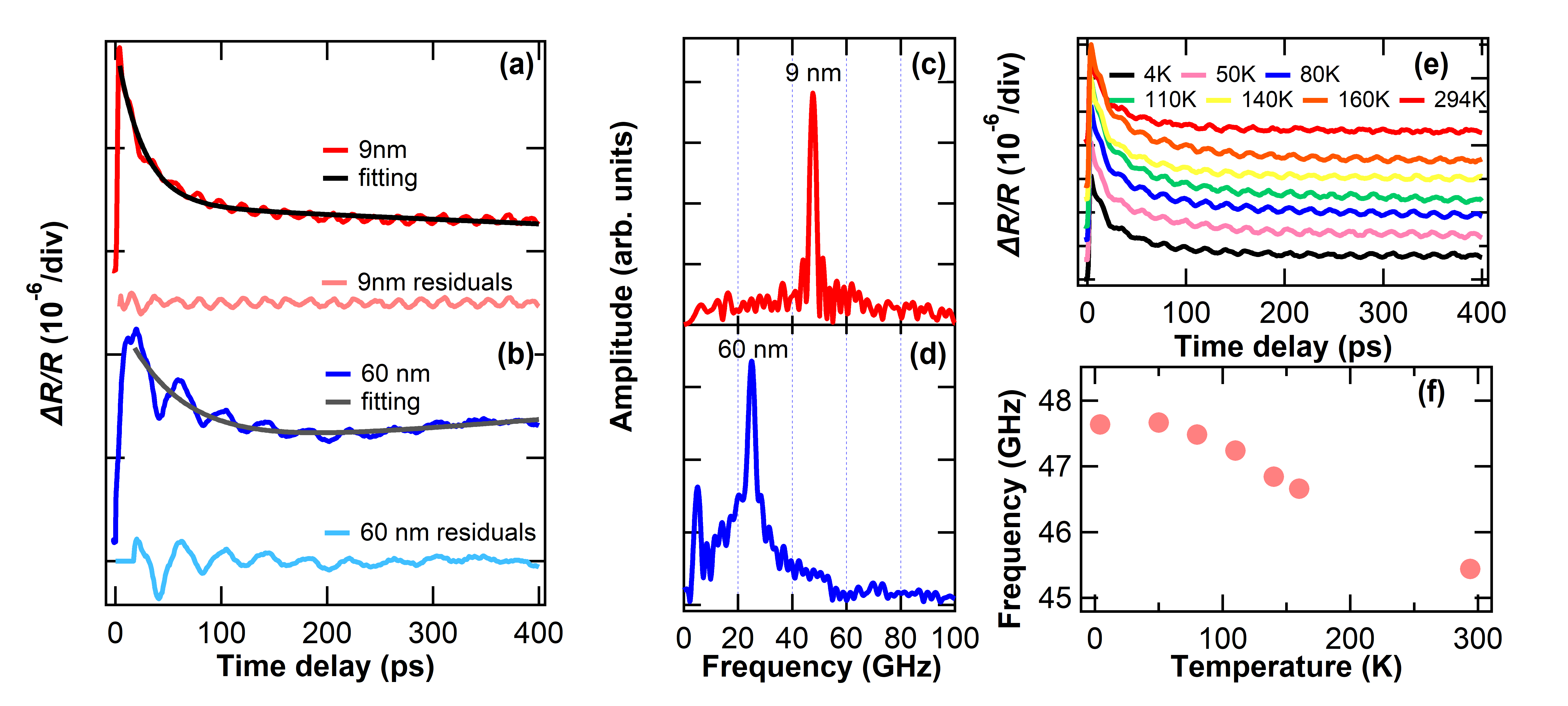}
	\caption{ Transient reflectivity change ${\Delta}R/R$ with corresponding biexponential fitting and residuals of (a) 9 nm and (b) 60 nm Ba-122 thin film at 5 K. Data are offset for clarity. Fourier spectra for the residuals of (c) 9 nm and (d) 60 nm Ba-122 thin film transient reflectivity. (e) Temperature-dependent transient reflectivity ${\Delta}R/R$ change (with offset) and (f) acoustic phonon frequency change for 9 nm Ba-122 thin film.}
	\label{fig2} 
\end{figure*}

To understand $u$ at different film thicknesses, we performed density functional theory (DFT) calculations \cite{Perdew} with a set of tested parameters \cite{CalPara}. Note that the sound propagates perpendicular to the film surface and thus mainly depends on the elastic properties along $c$-axis. On the other hand, the strain caused by the mismatch of substrate LiFe  ($a=4.027$ \AA,) and Ba-122 ($a=3.9612$ \AA)\cite{Song} mainly affects the $a$-$b$ plane, i.e., the tetragonal unit cell tends to be more expanded as it gets closer to the substrate, which is subject to intense distortion. The 9 nm film corresponds to 6 layers of the unit cell (Fig. 1b); while the 60 nm film corresponds to 40 layers, which is close to a bulk state. With this insight, we model the influence of film thickness by the extension of distortion in the $a$-$b$ plane (with tetragonal symmetry). In other words, we try to address the drastic change in $u$ by an elastic effect. Our purpose is to examine whether the elastic distortion can give rise to the significant changes observed in experiments.

The lattice parameter $a$ of substrate LiF sets an upper limit for distortion. That is, we examine $a$ in a range from 3.9612 {\AA} (bulk limit, 60 nm film) to 4.027 {\AA} (distortion limit, 9 nm film). Figure 3a illustrates the lattice parameter $c$ decreases with parameter $a$. Meanwhile, the cell volume $V$ is not constant, but will increase with $a$ (Fig. 3b). That suggests the bottom layer generally has a larger unit cell compared with the top (bulk). Elastic property, like Young's modulus, can be examined, which is expressed as \cite{Young}
\begin{equation}
Y=\frac{1}{V}\frac{{\partial}^2{E}}{\partial\varepsilon^2}.
\end{equation}
Volume $V$ is for the unit cell of the given distorted lattice $a$; $E$ is the total energy, $\varepsilon$ is the strain, which specially refers to $\varepsilon_{cc}$ in this scenario (the diagonal term along $c$-axis). With DFT and unit cell method, we can evaluate ${\partial}^2{E}/{\partial\varepsilon^2}$ by a fixed $a$ (maintain symmetry) and a slightly varied $c$ around the equilibrium $c_{eq}$ at given $a$.
\begin{equation}
\varepsilon_{cc}={\Delta}c/d,~{\Delta}c=c-c_{eq}(a)
\end{equation}
where $d$ is the thickness of thin film. The results are shown in Figure 3c and 3d. Notably, with increased $a$ and $V$, Young's modulus undergoes a four times increase (Fig. 3c). Combined with  
\begin{equation}
u=\sqrt{Y/{\rho}}=\sqrt{YV/m},
\end{equation}
we find that for bulk sample, the sound speed is 1240 m/s and for thin limit sample, the sound speed is 2416 m/s (Fig. 3e). Our theoretical calculations are in good agreement with the experimental results. 

Note that $\varepsilon$ is evaluated for the whole film, rather than for different locations within a film. Thus, the evaluated $Y$ physically means the average value for the whole thin film. It is important to realize one fitting curve (parabolic) in Fig. 3d will only correspond to one point in Fig. 3c. For example, in Fig. 3d, the red fitting curve corresponds to the bulk limit (red dashed line in Fig. 3c), and the points represent a series of small deviations around bulk; while the blue curve corresponds to the highly distorted limit (blue dashed lines in Fig. 3c).

\begin{figure*}[hbt!]
	\includegraphics[width=\textwidth]{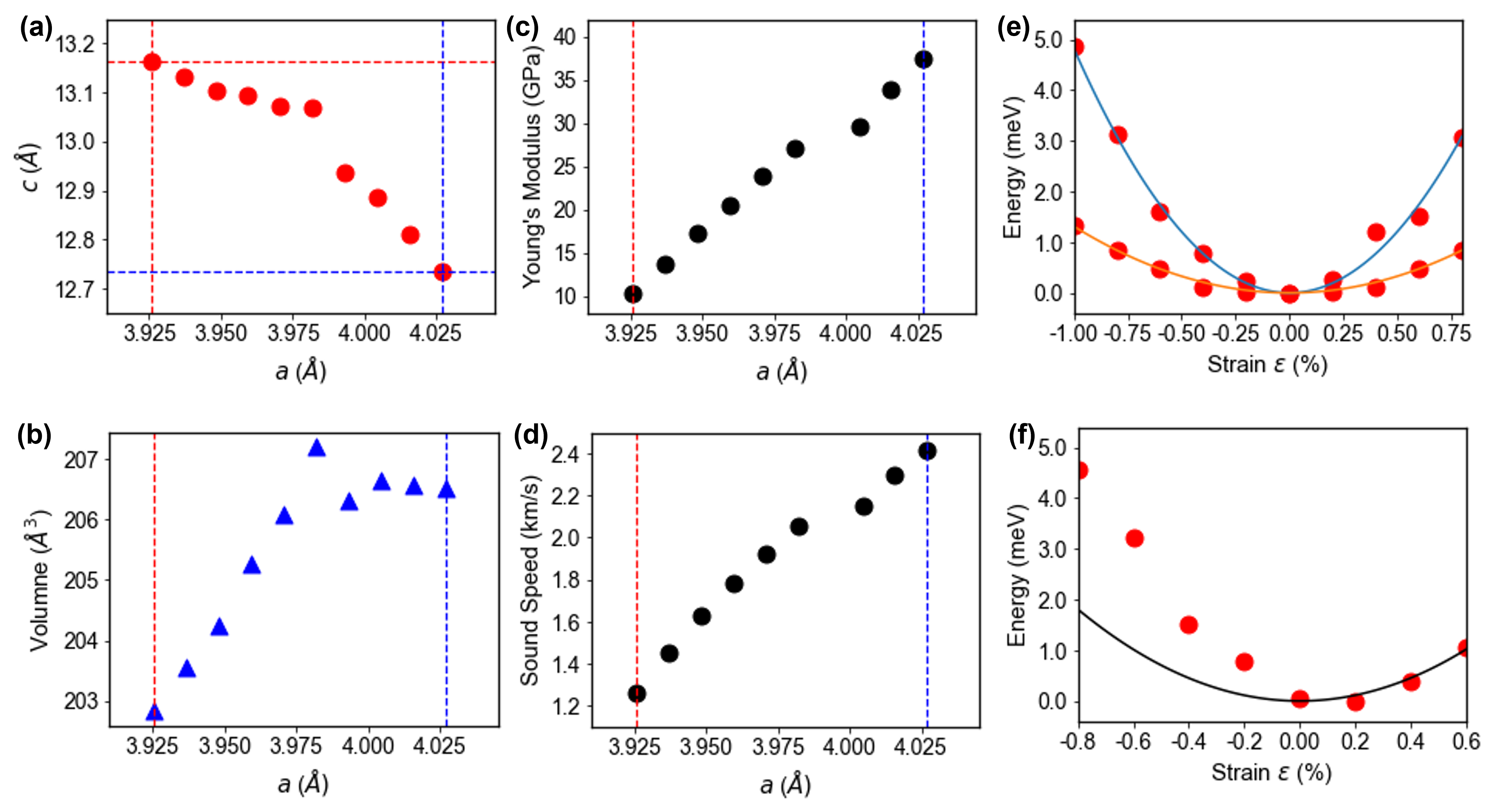}
	\caption{(a) Lattice parameter $c$ as a function of lattice parameter $a$. (b) Cell volumes as a function of lattice $a$. (c) Young's modulus dependence of $a$. (d) Elastic energy function E for 9 nm thin film (red dots, solid blue line is fitting) and bulk (red dots, solid red line is fitting). (e) Sound speeds as a function of lattice parameter \textit{a}. (f) Energy as a function of strain $\epsilon$ when $a$ = 3.99 \AA (red dots, solid black line is fitting).}
	\label{fig3} 
\end{figure*}

Noteworthy, discontinuity is observed when the lattice parameters are varied (Fig. 3a-c, e). This discontinuity corresponds to a phase transition, which is relevant to the double (or multiple) energy valleys\cite{Pbond}, which have been observed in layer-stacking structures, e.g., Ba-122/Fe based 1144-phase\cite{HC}. It occurs when $c$ parameter decreases, either for an applied pressure, or for an extended $a$-$b$ plane as done in this case. The $c$ lattice parameter will undergo a drastic change at a critical value, and it is called the collapse phase \cite{HC}. 
Consequently, from Fig. 3f, the strain energy function near the discontinuity is obviously deviant from parabolic line. That means $Y$ will exhibit quite different values at the two sides. 
It is interesting to check such abnormality using nano-scales imaging \cite{JW1} and other THz spectroscopy techniques \cite{XYang, JW2, JW3}.
In addition, although the mechanical properties are not the most intriguing aspects for intermetalic materials like Ba-122, the demonstrated ultrafast technique allows the sensitive measurement of $Y$ and $u$, which could provide a new avenue to obtain key information about the rich phase transition, electron-lattice coupling, etc. 

To summarize, we discovered propagating coherent acoustic phonon wavepackets in Ba-122 thin films using ultrafast transient reflectivity measurement. We reveal clear difference in frequency of the excited LA mode between 9 nm and 60 nm thin film sample. For the 9 nm thin film the LA frequency is nearly twice as large as that of the 60 nm samples. Temperature dependent LA mode frequency follows the structural phase transition of the sample. We attribute the change in mode frequency to the change in othorhombicity, which leads to the change in Young’s modulus, consistent with our calculations. Our results show that ultrafast transient reflectivity technique represents a powerful and sensitive tool to study the mechanical properties and the structural phase transition of ultrathin films with thickness even less than 10 nm.
Ultrast acoustic phonon detection may be extended to diverse complex thin films including superconducting \cite{S_1, S_2}, magnetic \cite{M_1} and photovoltaic materials \cite{P_1}.

\begin{acknowledgments}
This work was supported by National Science Foundation 1905981 (spectroscopy and analysis).
L.L. was supported by the Ames Laboratory, the US Department of Energy, Office of Science, Basic Energy Sciences, Materials Science and Engineering Division under contract No. DEAC0207CH11358 (technical assistance). The work at UW-Madison (synthesis and characterizations of epitaxial thin films) was supported by the US Department of Energy (DOE), Office of Science, Office of Basic Energy Sciences (BES), under award number DE-FG02-06ER46327 (C.B.E.).
Theory work at the University of Alabama, Birmingham (modeling) was supported by the US Department of Energy under contract \# DE-SC0019137 (M.M and I.E.P) and was made possible in part by a grant for high performance computing resources and technical support from the Alabama Supercomputer Authority.
\end{acknowledgments}



\begin{thebibliography}{99}
\bibitem{CalPara} All calculations of energy are based on density functional theory and Perdew Burke Ernzerhof (PBE) correlation functional, implemented by the Vienna ab initio simulation package (VASP). The projected augmented wave (PAW) pseudopotential method is employed. The plane wave energy truncation is 400 eV. The k-sampling has also been tested to ensure the convergence of energy. The magnetism is stripe AFM for Fe atoms. See Ref. \cite{Song, bqs} for a detailed description in various 122-phase compounds. Such magnetic structures have enlarged the unit cell, we have adopted the minimum sufficient $\sqrt{2}{\times}\sqrt{2}{\times}1$ 122-unit cell that contains 20 atoms.
\end{thebibliography}
\end{document}